\documentclass[a4paper,fleqn]{cas-sc}
\usepackage[authoryear]{natbib}
\usepackage{graphicx}
\usepackage{mathtools,bm}
\usepackage{siunitx}
\usepackage{setspace}

\DeclarePairedDelimiter{\abs}{\lvert}{\rvert}
\DeclarePairedDelimiter{\norm}{\lVert}{\rVert}

\newcommand*{\diff}{\mathrm{d}}

\begin{document}
\let\WriteBookmarks\relax
\shorttitle{Depth-averaged MPM}
\shortauthors{Kammholz et al.}

\title [mode = title]{A Highly Efficient Depth-Averaged Material Point Method for Geophysical Mass Flows}

\author[1,2,3]{Johann Kammholz}[orcid=0009-0006-4179-9770]
\credit{Writing -- original draft, Software, Formal analysis, Methodology, Investigation, Visualization}

\author[1,4]{Lars Blatny}[orcid=0000-0002-2016-7233]
\credit{Writing -- review \& editing, Software, Methodology, Supervision}

\author[1,2,3]{Yidong Zhao}[orcid=0000-0003-0703-8004]
\credit{Writing -- review \& editing, Validation, Software, Methodology}

\author[5]{Thomas Pauze}[orcid=0009-0004-5699-1031]
\credit{Software, Methodology}

\author[1,2,3]{Johan Gaume}[orcid=0000-0001-8931-752X]
\credit{Writing -- review \& editing, Supervision, Conceptualization, Funding acquisition}
\cormark[1]
\ead{jgaume@ethz.ch}

\address[1]{Institute for Geotechnical Engineering, Department of Civil, Environmental and Geomatic Engineering, ETH Zurich, Zurich, Switzerland}
\address[2]{WSL Institute for Snow and Avalanche Research SLF, Davos Dorf, Switzerland}
\address[3]{Climate Change, Extremes, and Natural Hazards in Alpine Regions Research Center CERC, Davos Dorf, Switzerland}
\address[4]{Institut Jean Le Rond d’Alembert, Sorbonne Université, CNRS - UMR 7190, Paris, France}
\address[5]{Météo-France, CNRS, Univ. Grenoble Alpes, Univ. Toulouse, CNRM, Centre d’Études de la Neige, Grenoble, France}

\cortext[cor1]{Corresponding author}

\begin{abstract}
Probabilistic hazard assessment of geophysical mass flows requires large ensemble simulations and thus fast and robust numerical models. We present an open-source depth-averaged material point method formulated in global Cartesian coordinates, using Lagrangian particles to track the flow margin. A separate basal friction step based on the Coulomb, Voellmy, or $\mu(I)$-rheology brings the flow to rest without artificial stopping criteria. Our model reproduces analytical solutions for a sliding block and dam-break problems and matches the deposit of a laboratory granular column collapse. A back-calculation of the 2023 Brienz/Brinzauls rockslide (Switzerland) at \SI{1}{\meter} resolution~(\num{190000} particles) runs faster than real time on a single CPU core. Calibrated to the observed runout, the three friction laws yield very similar results, but Coulomb friction shows a slightly wider affected area than Voellmy and $\mu(I)$. A probabilistic 1632-simulation ensemble at \SI{5}{\meter} resolution is computed in \SI{76}{\second}. For a single simulation at \SI{1}{\meter} resolution, the model is 69~times faster than RAMMS::AVALANCHE and 16~times faster than AvaFrame~com1DFA; for 1000~simulations at \SI{5}{\meter} resolution, it is 540 and 67~times faster, respectively, making catchment-style ensemble simulations feasible on a desktop computer.
\end{abstract}


\maketitle
\printcredits

\section{Introduction}

Gravity-driven geophysical mass flows, such as rockslides, debris flows, and snow avalanches, threaten mountain communities worldwide. These events frequently result in fatalities, cause damage to critical infrastructure, and incur economic costs related to hazard zoning, mitigation, and reconstruction~\citep{dowling2014debris, froude2018global}. Furthermore, climate change is altering the probability, intensity, and nature of these events, for example through permafrost degradation and changing precipitation patterns~\citep{gariano2016landslides, jacquemart2024detecting}. To address future climate uncertainty, reliable probabilistic hazard mapping requires running ensemble simulations across wide parameter ranges~\citep{buhler2022automated, ortner2025novel}, necessitating computationally efficient models.

The modeling of geophysical mass flows has evolved significantly over the past decades. Early efforts relied on empirical methods based on topographic parameters~\citep{lied1980empirical} and analytical calculations~\citep{salm1990berechnung} that have since been refined into modern Geographic Information Systems~(GIS)-based tools~\citep{horton2013flow, berti2014dflowz}. However, while practical and widely used for preliminary assessments, these approaches lack a dynamic representation of the flow physics. The foundational basis for modern continuum models was established when \citet{savage1989motion} extended the Saint--Venant shallow-water equations to dry granular flows. The Saint--Venant equations, which are derived by integrating the incompressible Navier--Stokes equations over the flow depth, are written in Cartesian coordinates without slope corrections and are only valid for small slope angles (almost flat topography), rendering them inapplicable to geophysical mass flows. The Savage--Hutter model overcomes this limitation by integrating the balance laws in coordinates aligned with the bed, thereby only requiring small slope variations, i.e., small curvature~\citep{savage1991dynamics, pudasaini2007avalanche}. While strictly one-dimensional, this foundational model combines a Coulomb basal friction law with an earth-pressure closure for internal stresses. Earlier and independently, similar depth-averaged avalanche models were pioneered in the Soviet Union~\citep{grigorian1967new, eglit2020review}. At the time, solving these equations numerically in two dimensions was challenging. Although first two-dimensional formulations existed~\citep{kulikovskii1973two, hutter1993two, gray1999gravity}, the first operational numerical tools were restricted to one-dimensional depth-averaged models, such as DAN~\citep{hungr1995model} and AVAL-1D~\citep{bartelt1999calculating}, which provided insights into flow velocities and flow heights along predefined one-dimensional profiles. As the theoretical framework, numerical schemes, and computational power advanced, a wide variety of two-dimensional depth-averaged models were developed~\citep{pastor2015depth}. They can be clustered by their numerical schemes: Eulerian models based on finite volume methods~(FVM) or finite difference methods~(FDM), e.g., SHALTOP~\citep{mangeney2003numerical, bouchut2004gravity}, TITAN2D~\citep{patra2005parallel}, RAMMS~\citep{christen2010ramms}, D-Claw~\citep{iverson2014depth, george2014depth}, and r.avaflow~\citep{pudasaini2012general, mergili2017r}, and Lagrangian models based on smoothed-particle hydrodynamics~(SPH), e.g., DAN3D~\citep{mcdougall2004model, hungr2009two}, GeoFlow-SPH~\citep{pastor2009depth, pastor2014application}, and SAMOS-AT~\citep{sampl2009avalanche}, which served as the basis of the AvaFrame com1DFA module~\citep{tonnel2023avaframe}.

Despite these advancements, significant limitations remain. Eulerian models struggle to sharply track the exact flow margin and suffer from numerical diffusion~\citep{leveque2002finite}. Furthermore, they typically require complex shock-capturing numerical techniques to resolve discontinuities~\citep{nessyahu1990non, gray2003shock}, and flow height cut-off thresholds or height-positivity preserving schemes~\citep{mangeney2003numerical}. Additionally, some models require artificial stopping criteria to handle material deposition~\citep{christen2010ramms, rauter2016snow, goodwin2024has}. Conversely, while Lagrangian SPH models naturally track the flow margin, they often rely on artificial viscosity or smoothing to stabilize shocks and suppress oscillations, and require complex neighbor search algorithms~\citep{monaghan2005smoothed}. Three-dimensional depth-resolved models, such as 3D MPM~\citep{cicoira2022towards, blatny2025matter}, 3D SPH~\citep{nikooei2020studying}, or 3D FVM~\citep{boetticher2016debrisintermixing}, are powerful tools for fundamental process understanding, detailed hazard assessment, and mitigation structure design. Due to their high computational cost, however, these models are generally applied to individual slopes or selected high-risk flow paths rather than catchment-scale probabilistic hazard mapping requiring large ensembles of simulations. Finally, several established models (2D depth-averaged and 3D depth-resolved) remain commercial or closed-source.

To address these gaps, this paper presents a depth-averaged material point method for rapid granular mass flows. The material point method~\citep[MPM,][]{sulsky1994particle} is a hybrid Eulerian--Lagrangian continuum method, and can be considered an extension of the particle-in-cell~(PIC) and fluid-implicit-particle~(FLIP) schemes. The material domain is discretized into moving Lagrangian material points that carry the state variables, while a stationary Eulerian grid is used to compute spatial derivatives and solve the equations of motion. Quantities are mapped between material points and grid nodes using interpolation functions. The particles track the flow margins without numerical diffusion, and the forces are computed on the grid with fixed interpolation stencils, which requires no neighbor search. MPM has its own numerical issues. Pure PIC transfers are strongly dissipative and pure FLIP transfers are prone to noise, which is why blended or higher-order transfers are used~\citep{jiang2015affine}. Grid nodes that receive only a small fraction of a particle's mass can obtain spurious accelerations, and linear interpolation functions cause cell-crossing errors~\citep{steffen2008analysis}. Therefore, quadratic or cubic B-splines are used~\citep{stomakhin2013material, gaume2018dynamic}. To date, only a few works on depth-averaged MPM have been published. \citet{abe2007new} and~\citet{abe2016numerical} were the first to propose the depth-averaged MPM scheme. \citet{guillet2023depth} extended this method to model snow avalanche release using a cohesive Cam-Clay model for the slab and an elastic-quasi-brittle weak layer. \citet{fois2024semi} adapted it for landslides using a Bingham model coupled with Voellmy basal friction. Additionally, \citet{huth2021generalized} applied a generalized interpolation material point method~(GIMP) to shallow ice shelves.

Our depth-averaged MPM framework introduces several geoscientific and computational improvements. First, while the Savage--Hutter model relies on bed-aligned curvilinear coordinates, we adopt a global Cartesian coordinate system similar to the original Saint--Venant equations. This choice is motivated by mathematical simplicity and naturally aligns with vertically measured remote sensing data~\citep{denlinger2004granular}. Second, we split each time step into an explicit update without friction and a semi-implicit friction step~\citep{mangeney2003numerical}. This allows the flow to stop naturally without the need for an artificial stopping criterion. Third, the model integrates the shear-rate-dependent $\mu(I)$-rheology alongside Coulomb and Voellmy basal resistance laws. Fourth, our model is 16 to 540~times faster than the established operational models RAMMS and AvaFrame com1DFA.

We verify the model against analytical solutions, validate it against a laboratory granular column collapse, and apply it to the 2023 Brienz/Brinzauls rockslide (Switzerland), with release heights taken from differenced drone surveys. A back-calculation with Coulomb, Voellmy, and $\mu(I)$ friction shows that deposit data alone cannot discriminate between friction laws. An ensemble of 1632~simulations yields an inundation probability map, demonstrating that probabilistic hazard assessment is feasible on a desktop computer.

\section{Methods}\label{sec:methods}

\subsection{Governing Equations}

\subsubsection{Depth-Averaged Conservation Equations}

Our model relies on a continuum mechanical description of granular flows under the following assumptions:

\begin{enumerate}
    \item The material is incompressible, and there is no entrainment.
    \item The flowing mass is shallow, i.e., the flow height~$h$ varies over a characteristic horizontal length~$L$ that is much larger than the characteristic flow height~$H$ (aspect ratio $\epsilon=\frac HL \ll 1$). The flow velocity is strictly parallel to the bed, and the velocity profile is uniform over the flow height.
    \item Bed-normal accelerations (and hence curvature effects) are neglected. The pressure balances the bed-normal component of the weight.
    \item The internal stress state is isotropic. The free surface is stress-free, while the bed provides resistance defined by a basal friction law.
\end{enumerate}

We use a global Cartesian coordinate system where the $x$- and $y$-coordinates are horizontal, and the $z$-coordinate is vertical (opposite to the direction of gravity). As in the formulation by \citet{hergarten2015modelling}, the flow velocity is averaged vertically and measured horizontally, and the flow height is measured vertically.

Depth-integrating the 3D mass conservation equation over the flow height yields
\begin{equation}
    \frac{\partial h}{\partial t} + \nabla \cdot (h \bm v) = 0,
\end{equation}
where $h(x,y,t) = s(x,y,t) - b(x,y)$ is the vertical flow height (with the free surface~$s$ and the bed elevation~$b$), $t$ is time, $\bm v = (v_x, v_y)^\mathrm T$ is the depth-averaged horizontal velocity vector, and $\nabla = (\frac{\partial}{\partial x}, \frac{\partial}{\partial y})^\mathrm T$.

Similarly, depth-integrating the horizontal components of the 3D momentum conservation equation yields
\begin{equation}\label{eq:momentumBalance}
    \rho h \frac{\diff\bm v}{\diff t} = \nabla \cdot (h\bm\sigma) + \rho h \bm g_\parallel + \frac{\bm\tau_b}{\cos\theta},
\end{equation}
where $\rho$ is the density, $\frac{\diff}{\diff t}$ is the material derivative, and $\bm\sigma$ is the depth-averaged Cauchy stress tensor (compression is negative). The term $\rho h \bm g_\parallel$ is the horizontal component of the gravitational driving force per unit horizontal area, resulting from the normal stress exerted by the bed on the flowing mass, where $\bm g_\parallel = -g\cos^2\theta\,\nabla b$ is the horizontal component of the bed-parallel gravitational acceleration, $g$ is gravity, and $\theta$ is the bed inclination angle satisfying $\tan\theta = \norm{\nabla b}$. Finally, $\bm\tau_b$ denotes the horizontal components of the basal frictional traction acting opposite to~$\bm v$. The factor $\frac{1}{\cos\theta}$ projects the traction from a force per unit bed area to a force per unit horizontal area.

In this two-dimensional formulation, we solve for the flow height~$h$ and the horizontal velocity~$\bm v$. Given the horizontal velocity, the bed topography~$b(x,y)$, and the assumption that the flow is parallel to the bed, the vertical velocity~$v_z$ of a material parcel at the vertical position~$z$ can be obtained by defining the bed-outward normal vector
\begin{equation}
    \bm n = \frac{1}{\sqrt{1 + \norm{\nabla b}^2}} \begin{pmatrix} -\nabla b \\ 1 \end{pmatrix} = \cos\theta \begin{pmatrix} -\nabla b \\ 1 \end{pmatrix},
\end{equation}
and imposing the boundary condition $\bm v_\text{3D} \cdot \bm n = 0$. The full three-dimensional velocity vector is then $\bm v_\text{3D} = \left(v_x, v_y, \nabla b \cdot \bm v\right)^\mathrm T$.

\subsubsection{Stress Closure}

Following Assumption~4, the stress state is isotropic: the normal stress takes the same value~$\sigma_n$ on any plane through a point, and shear stresses vanish. In particular, the bed-normal, vertical, and lateral normal stresses coincide, which corresponds to an earth pressure coefficient~$K = 1$. Before depth-averaging, $\sigma_n$ is assumed to be hydrostatic,
\begin{equation}
    \sigma_n(z) = -\rho g (s - z) \cos^2\theta, \qquad b \le z \le s,
\end{equation}
where one factor of $\cos\theta$ comes from the bed-normal component of gravity, and the other from converting the vertical depth $s - z$ into a bed-normal depth. Depth-averaging yields the depth-averaged normal stress
\begin{equation}\label{eq:sigma_n_bar}
    \bar\sigma_n = \frac1h\int_b^{s} \sigma_n(z)\,\mathrm dz = -\frac{\rho g h \cos^2\theta}{2},
\end{equation}
and the depth-averaged stress tensor is $\bm\sigma = \bar\sigma_n\bm I$.

This assumption effectively treats the granular mass as a fluid without internal friction~\citep{gray1999gravity} and is the usual stress closure in operational models for hazard assessment of geophysical mass flows~\citep{christen2010ramms, sampl2009avalanche}.

\subsubsection{Basal Friction}\label{sec:basal_friction}

Basal friction governs energy dissipation at the flow base. We implement three basal resistance laws as alternative closures for the basal traction~$\bm\tau_b = (\tau_{xz}, \tau_{yz})^\mathrm T$.

\paragraph{Coulomb}
The first option is Coulomb friction,
\begin{equation}
    \bm\tau_b = \mu\sigma_n(b)\frac{\bm v}{\norm{\bm v_\text{3D}}},
\end{equation}
where $\mu = \tan\delta$ is the Coulomb friction coefficient (with the basal friction angle~$\delta$), and the normal stress at the bed is
\begin{equation}
    \sigma_n(b) = -\rho g h \cos^2\theta.
\end{equation}

\paragraph{Voellmy}
The second option is the model by \citet{voellmy1955zerstorungskraft},
\begin{equation}
    \bm\tau_b = \left(\mu\sigma_n(b) - \frac{\rho g \norm{\bm v_\text{3D}}^2}{\xi}\right)\frac{\bm v}{\norm{\bm v_\text{3D}}}.
\end{equation}
This formulation extends the Coulomb friction model by adding a velocity-dependent turbulent term governed by the parameter~$\xi$. With this model, a steady-state velocity exists for $\theta > \arctan \mu$, and its magnitude in bed-parallel direction can be derived as
\begin{equation}
    v_\infty = \sqrt{(h \cos\theta) \xi (\sin\theta - \mu \cos\theta)}.
\end{equation}
Originally employed for snow avalanches, the Voellmy model is widely popular in hazard assessment of geophysical mass flows and is implemented in various operational models~\citep{christen2010ramms, mcdougall2004model, tonnel2023avaframe}. The model generally provides satisfactory results for avalanches and debris flows, allowing for calibration of both runout distance and flow velocity.

\paragraph{Granular Rheology}
The third option is the $\mu(I)$-rheology~\citep{jop2006constitutive},
\begin{equation}
    \bm\tau_b = \mu(I)\sigma_n(b)\frac{\bm v}{\norm{\bm v_\text{3D}}}.
\end{equation}
The friction coefficient $\mu(I)$ depends on the inertial number~$I = \frac{\dot\gamma d}{\sqrt{P/\rho_s}}$, where $\dot\gamma$ is the shear rate, $P = -\sigma_n(z)$ the local confining pressure, $d$ the grain diameter, and $\rho_s = \frac{\rho}{\phi}$ the grain density, with $\phi$ the solid volume fraction. It typically follows the empirical law
\begin{equation}\label{eq:mu_I}
    \mu(I) = \mu_1 + \frac{\mu_2-\mu_1}{\frac{I_0}{I}+1},
\end{equation}
where $\mu_1$ and $\mu_2$ are the limiting friction coefficients for $I\to0$ (quasi-static) and $I\to\infty$ (dynamic), respectively, and $I_0$ is a material constant controlling the transition between them. In steady, uniform flow down an incline, assuming a hydrostatic pressure distribution, the ratio of shear stress to pressure equals $\tan\theta$ at every depth, i.e., $\mu(I)=\tan\theta$, so $I$ is uniform across the flow height and the velocity follows a Bagnold profile. Depth-averaging this profile relates $I$ to the depth-averaged velocity~\citep{gray2014depth},
\begin{equation}\label{eq:inertial_number_da}
    I = \frac52 \frac{d \norm{\bm v_\text{3D}}}{h \cos^2\theta\,\sqrt{g h \phi}}.
\end{equation}
Inverting Eq.~\eqref{eq:inertial_number_da} yields the steady-state bed-parallel velocity
\begin{equation}
    v_\infty = \frac{2I}{5d} h \cos^2\theta\,\sqrt{g h \phi},
\end{equation}
with $I$ calculated from Eq.~\eqref{eq:mu_I} using $\mu(I) = \tan\theta$. Such a steady state exists only for $\mu_1 < \tan\theta < \mu_2$. Unlike the Voellmy parameters, the $\mu(I)$ parameters can be calibrated in laboratory experiments, and the framework can successfully capture the complex flow behavior of dense granular flows~\citep{gdr2004dense,blatny2024critical}.

In the numerical scheme, the friction term of Eq.~\eqref{eq:momentumBalance} is not included in the nodal forces but applied in a separate friction step. This prevents non-physical velocity reversal and allows the flow to come to rest.

\subsection{Depth-Averaged Material Point Method}

We solve the depth-averaged momentum conservation equation, Eq.~\eqref{eq:momentumBalance}, with MPM. In the depth-averaged framework, introduced by \citet{abe2016numerical}, the material point representation is modified compared to 3D depth-resolved MPM. Rather than 3D points, the material points (hereafter referred to as particles) represent vertical columns with a variable flow height. Throughout this section, the index~$p$ denotes particle quantities and the index~$i$ grid quantities.

\begin{figure}[pos=htbp]
    \centering
    \includegraphics[width=120mm]{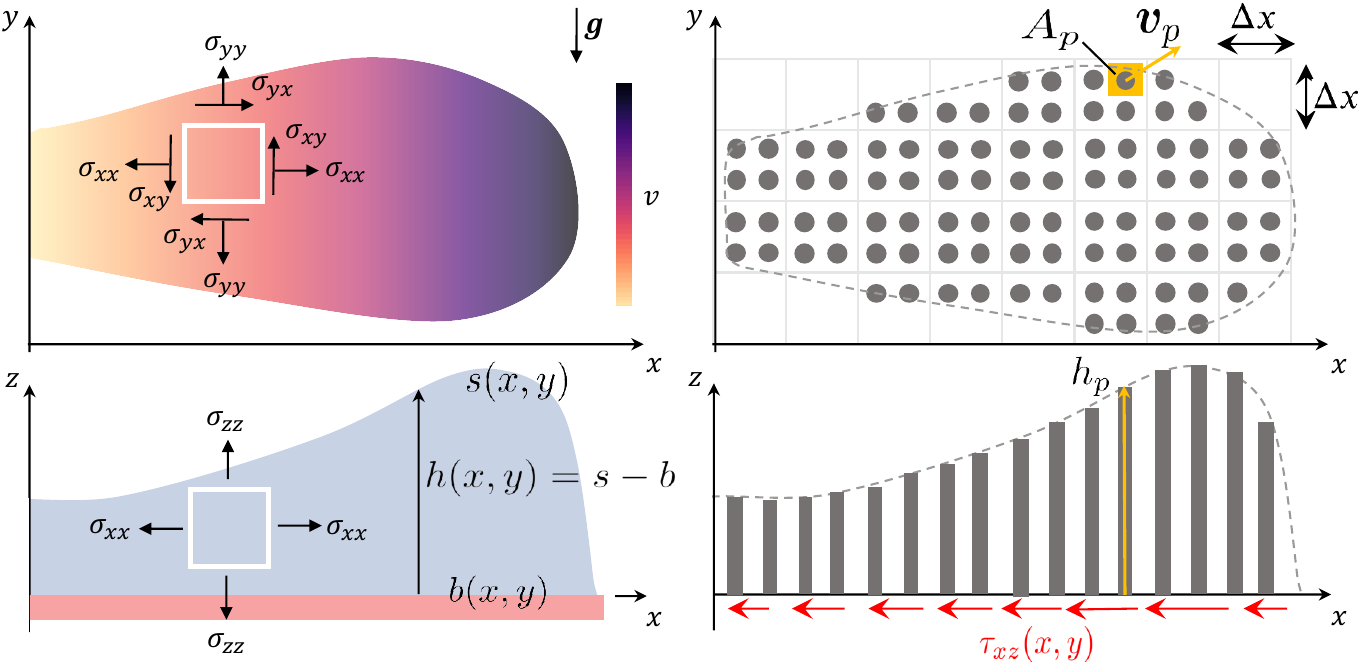}
    \caption{Left: continuum view of a geophysical mass flow. Right: discretized view in the framework of depth-averaged MPM. Top: view from above showing the horizontal plane. Bottom: view from the side showing a cross-section of the flow. Note that the slope angle $\theta$ is zero in this figure.}
    \label{fig:2_2_dampm}
\end{figure}

\paragraph{Spatial Discretization and Input/Output}
The domain is discretized by a regular Cartesian grid with spacing~$\Delta x = \Delta y$. The topography is defined on the grid using a digital elevation model~(DEM) imported from a GIS raster file, which provides the bed elevation~$b(\bm x_i)$ and dictates the grid resolution~$\Delta x$. The grid is also used to compute the topographic gradient~$\nabla b$. The flowing mass is discretized into particles, each associated with a horizontal position~$\bm x_p = (x_p, y_p)^\mathrm T$, horizontal velocity~$\bm v_p$, vertical flow height~$h_p$, planform area~$A_p$, and density~$\rho_p$. A schematic is shown in Fig.~\ref{fig:2_2_dampm}. The mass of particle~$p$ is
\begin{equation}
    m_p = \rho_p h_p A_p,
\end{equation}
and its vertical position is $z_p = b(\bm x_p)$. The release area can be imported from a polygon shapefile or a GIS raster file, the latter accommodating non-uniform release heights. Simulation outputs are exported as GIS raster files for Eulerian grid data and PLY~(Polygon File Format) point clouds for Lagrangian particle data.

\paragraph{Initial Sampling} Particle sampling follows either regular Cartesian sampling or Poisson disk sampling~\citep{bridson2007fast}. While regular sampling is computationally trivial, its periodic structure introduces artificial directional bias as the spacing between two particles strongly depends on the direction. Therefore, we recommend using Poisson disk sampling, which enforces a minimum inter-particle distance and yields an isotropic, unbiased coverage. We typically sample four particles per cell.

\paragraph{Algorithm} One time step from $t^k$ to $t^{k+1}$ consists of the following steps:

\begin{enumerate}
    \item \textbf{Update time step size:} The time step size~$\Delta t^k$ is constrained by the Courant--Friedrichs--Lewy~(CFL) condition
    \begin{equation}
        \Delta t^k = \min\left(\frac{C_\text{cfl}\Delta x}{\max_p \norm{\bm v_p^k}}, \Delta t_\text{max}\right),
    \end{equation}
    where $C_\text{cfl}\in(0,1]$ is the Courant number and $\Delta t_\text{max}$ is the maximum time step size allowed. We typically choose $C_\text{cfl} = 0.5$. In the PIC/FLIP scheme used here, artificially reducing $\Delta t^k$ below the necessary stability bound amplifies numerical dissipation by increasing the frequency of particle--grid transfers~\citep{jiang2015affine}.

    \item \textbf{Resize grid:} The grid is extended or contracted to align with the bounding box that contains all particles, including a buffer to account for the support of the interpolation functions.

    \item \textbf{Particle sorting:} Every 50 time steps, particles are sorted in memory based on their current physical position. This ensures that spatially adjacent particles are stored in contiguous memory locations, reducing random memory accesses and cache misses~\citep{gao2018gpu}.

    \item \textbf{Particle-to-Grid~(P2G):} We use quadratic B-spline shape functions for their lower cost compared to cubic B-splines (9 instead of 16 nodes per particle):
    \begin{equation}
        N_i(\bm x_p^k) = N\!\left(\frac{x_p^k-x_i}{\Delta x}\right)N\!\left(\frac{y_p^k-y_i}{\Delta y}\right),
    \end{equation}
    where
    \begin{equation}
        N(u) = \begin{cases}
            \frac34 - u^2 &\text{if}\ \abs{u} < \frac12,\\
            \frac12(\frac32-\abs{u})^2 &\text{if}\ \frac12 \leq \abs{u} < \frac32,\\
            0 &\text{else.}
        \end{cases}
    \end{equation}
    Mass and momentum are transferred from the particles to the grid, where grid mass is
    \begin{equation}\label{eq:m_grid}
        m_i^k = \sum_p m_p N_i(\bm x_p^k),
    \end{equation}
    and grid velocity is
    \begin{equation}
        \bm v_i^k = \frac{1}{m_i^k} \sum_p m_p \bm v_p^k N_i(\bm x_p^k).
    \end{equation}
    The same interpolation is used for grid flow height
    \begin{equation}
        h_i^k = \frac{1}{m_i^k} \sum_p m_ph_p^kN_i(\bm x_p^k),
    \end{equation}
    which is used for GIS raster data output and in the Voellmy and $\mu(I)$ friction laws.

    The discretized weak form of Eq.~\eqref{eq:momentumBalance} gives the internal and external forces
    \begin{align}
        \bm F_i^{\text{int},k} &= - \sum_p h_p^k A_p^k \, \bm\sigma_p^k \cdot \nabla N_i(\bm x_p^k), \label{eq:F_int}\\
        \bm F_i^{\text{ext},k} &= - \sum_p m_p g \cos^2\theta_p\,\nabla b(\bm x_p^k) N_i(\bm x_p^k), \label{eq:F_ext}
    \end{align}
    with the particle stress~$\bm\sigma_p^k = -\frac12 \rho_p g h_p^k \cos^2\theta_p\,\bm I$, particle terrain angle~$\theta_p$, and particle terrain gradient~$\nabla b(\bm x_p^k)$.

    \item \textbf{Update grid velocities:} Grid acceleration is computed as
    \begin{equation}\label{eq:a_grid}
        \bm a_i^k = \frac{\bm F_i^{\text{int},k} + \bm F_i^{\text{ext},k}}{m_i^k},
    \end{equation}
    and grid velocity is updated as
    \begin{equation}\label{eq:v_trial}
        \bm v_i^* = \bm v_i^k + \Delta t^k \bm a_i^k.
    \end{equation}
    Here, $\bm v_i^{*}$ is an intermediate trial velocity that does not yet include basal friction. The 3D intermediate velocity~$\bm v_{i,\text{3D}}^*$ is recovered from the no-penetration boundary condition.

    \item \textbf{Friction step:} Basal friction reduces the magnitude of the trial velocity~$\bm v_i^{*}$ without changing its direction. Using the grid terrain angle~$\theta_i$, we determine the frictional deceleration magnitude
    \begin{equation}
        a_\text{fric}^k = \mu_\text{eff} g \cos\theta_i,
    \end{equation}
    where the expression for $\mu_\text{eff}$ depends on the selected basal resistance law, e.g., as a function of $\bm v_i^*$ for the rate-dependent laws. The velocity is then updated according to
    \begin{equation}
        \bm v_i^{k+1} =
        \begin{cases}
            \bm v_i^{*} \left( 1 - \dfrac{\Delta t^k a_\text{fric}^k}{\norm{\bm v_{i,\text{3D}}^*}} \right),
            &\text{if}\ \Delta t^k a_\text{fric}^k < \norm{\bm v_{i,\text{3D}}^*}, \\
            \bm 0, &\text{else}.
        \end{cases}
    \end{equation}

    \item \textbf{Grid-to-Particle~(G2P):} The updated grid velocities are transferred back to the particles using a blended PIC/FLIP scheme
    \begin{equation}
        \bm v_p^{k+1} = \left( \bm v_p^k + \sum_i (\bm v_i^{k+1} - \bm v_i^k) N_i(\bm x_p^k) \right)\alpha + \left( \sum_i \bm v_i^{k+1} N_i(\bm x_p^k) \right)(1 - \alpha),
    \end{equation}
    where $\alpha\in[0,1]$ controls the linear combination of PIC and FLIP. We typically choose $\alpha$ between 0.9 and 0.95, which is a standard choice in MPM to suppress high-frequency noise while preserving energy~\citep{stomakhin2013material}.

    \item \textbf{Update particles:} Particle positions are updated as
    \begin{equation}
        \bm x_p^{k+1} = \bm x_p^k + \Delta t^k \sum_i \bm v_i^{k+1} N_i(\bm x_p^k).
    \end{equation}
    The trace of the strain increment is computed as
    \begin{equation}
        \mathrm{tr}(\Delta\bm\varepsilon_p^k) = \Delta t^k \sum_i \bm v_i^{k+1}\cdot\nabla N_i(\bm x_p^k).
    \end{equation}
    The flow height update follows from mass/volume conservation as
    \begin{equation}
        h_p^{k+1} = \frac{h_p^k}{1 + \mathrm{tr}(\Delta\bm\varepsilon_p^k)},
    \end{equation}
    and the particle area is accordingly
    \begin{equation}
        A_p^{k+1} = \frac{m_p}{h_p^{k+1}\rho_p}.
    \end{equation}
\end{enumerate}

\section{Results}\label{sec:results}

Unless stated otherwise, all simulations use $g = \SI{9.8045}{\meter\per\second\squared}$, quadratic B-splines, Poisson disk sampling, $C_\text{cfl} = 0.5$, and $\alpha = 0.9$. Hereafter, we refer to the presented depth-averaged MPM as DAMPM in equations and figures.

\subsection{Verification}

\subsubsection{Constant Slope}

We first consider a rigid block that slides down an inclined plane from rest. The block has a bed-normal height~$h_n = \SI{2}{\meter}$, i.e., a vertical height~$h = \frac{h_n}{\cos\theta}$, and is discretized by a single particle. As the block is rigid, internal forces are disabled, $\bm F_i^\text{int} = \bm 0$ in Eq.~\eqref{eq:F_int}, so that $h$ remains constant. We use $\Delta x = \SI{1}{\meter}$ and a maximum time step size~$\Delta t_\text{max} = \SI{0.04}{\second}$, and test all three basal friction laws: Coulomb with $\mu = 0.2$; Voellmy with $\mu = 0.2$ and $\xi = \SI{2000}{\meter\per\second\squared}$; and $\mu(I)$ with $\mu_1 = 0.2$, $\mu_2 = 0.65$, $I_0 = 0.25$, $d = \SI{0.1}{\meter}$, and $\phi = 0.6$. The reference solution follows from the equation of motion in bed-parallel direction that we integrate with the Runge--Kutta method RK45 implemented in \texttt{scipy.\allowbreak integrate.\allowbreak solve\_ivp} using tight tolerances. Horizontal positions and velocities are converted after the simulation to bed-parallel coordinates $x_\parallel = \frac{x}{\cos\theta}$ and $v_\parallel = \frac{v_x}{\cos\theta}$.

\begin{figure}[pos=htbp]
    \centering
    \includegraphics{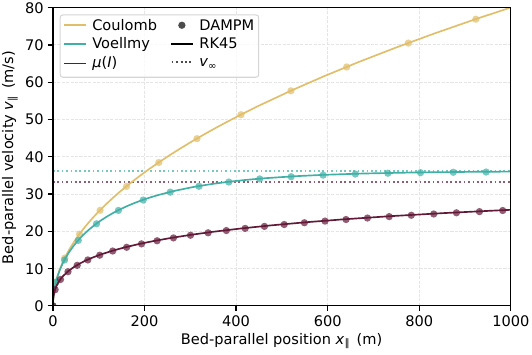}
    \caption{Transient velocity of a rigid block sliding from rest down an inclined plane at $\theta = \ang{30}$. Markers: depth-averaged MPM results using a single particle, plotted at intervals of \SI{2}{\second}. Lines: reference solution, integrated with the Runge--Kutta method RK45. Dotted lines: analytical steady-state velocities $v_\infty$ (no steady state exists for Coulomb). Note that the analytical solution for $\mu(I)$ approaches $v_\infty$ only beyond the plotted range after several kilometers. This is because $\theta$ is very close to $\arctan(\mu_2)$.}
    \label{fig:3_1_1_velocity_vs_position}
\end{figure}

To illustrate the transient acceleration of the block, we first perform one simulation per friction law at a fixed slope angle~$\theta = \ang{30}$. In all cases, the simulation agrees with the reference solution without visible deviation~(Fig.~\ref{fig:3_1_1_velocity_vs_position}). With Coulomb friction, the block accelerates indefinitely. With the velocity-dependent laws, velocity correctly approaches the analytical steady-state velocity~$v_\infty$~(Sec.~\ref{sec:basal_friction}).

\begin{figure}[pos=htbp]
    \centering
    \includegraphics{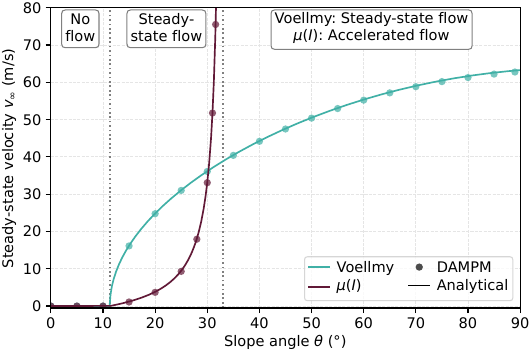}
    \caption{Steady-state velocity $v_\infty$ (in bed-parallel direction) of a rigid block versus slope angle $\theta$. Markers: depth-averaged MPM results extracted from individual simulations (run to sufficiently large $t$) at varying slope angles. Lines: analytical solutions~(Sec.~\ref{sec:basal_friction}). For $\theta > \ang{70}$, the time step size is too large and the model slightly underestimates $v_\infty$. A smaller time step size recovers $v_\infty$ more accurately. The last marker is at $\theta = \ang{89}$, as $\theta = \ang{90}$ cannot be represented in the horizontal coordinate system.}
    \label{fig:3_1_1_steady-state_velocity}
\end{figure}

To systematically evaluate the model's accuracy across inclinations, we conduct a parameter sweep over the slope angle~$\theta$ for the Voellmy and $\mu(I)$ friction laws. Each simulation is run until the block reaches its steady-state velocity. Figure~\ref{fig:3_1_1_steady-state_velocity} compares these velocities against the analytical solutions. The model captures the transition from the no-flow regime (where $\theta$ is too shallow to overcome static friction) to the flowing regime, and agrees with the theoretical curves across the angular range.

\subsubsection{Dam-Break Problem}

The dam-break problem tests the pressure gradient term $\nabla\cdot(h\bm\sigma)$ and the resulting wave speed. A mass is released instantaneously over a dry bed with the initial condition
\begin{equation}\label{eq:dam_break_ic}
\begin{split}
    h_n(x_\parallel, t_0) &= \begin{cases} h_0, & x_\parallel \leq 0,\\ 0, & x_\parallel > 0, \end{cases} \\
    v_\parallel(x_\parallel, t_0) &= 0,
\end{split}
\end{equation}
with $t_0=\SI{0}{\second}$ and $h_0 = \SI{10}{\meter}$. The setup replicates that of \citet{mcdougall2004model}. The reference is the analytical solution of \citet{mangeney2000analytical} for an infinitely long reservoir on an inclined plane with Coulomb friction~($\delta < \theta$), which includes the flat, frictionless case~\citep{ritter1892fortpflanzung}. We solve this 1D problem in 2D with an initial reservoir of \SI{1000}{\meter} $\times$ \SI{200}{\meter}, in which the cross-slope~($y$)~forces in Eqs.~\eqref{eq:F_int} and~\eqref{eq:F_ext} are set to zero. The upstream end at $x_\parallel = \SI{-1000}{\meter}$ is not treated explicitly, i.e., the dam also collapses upstream. In the simulation, the initial discontinuity at $x_\parallel = \SI{0}{\meter}$ is smoothed over $3\Delta x$. We use cubic B-splines, $\Delta x = \SI{1}{\meter}$, pure FLIP~($\alpha = 1$), $C_\text{cfl} = 0.1$, and $\Delta t_\text{max} = \SI{0.04}{\second}$. To extract one-dimensional profiles, particle flow heights and velocities are averaged over $y$ in bins along $x_\parallel$.

\begin{figure}[pos=htbp]
    \centering
    \includegraphics[width=\linewidth]{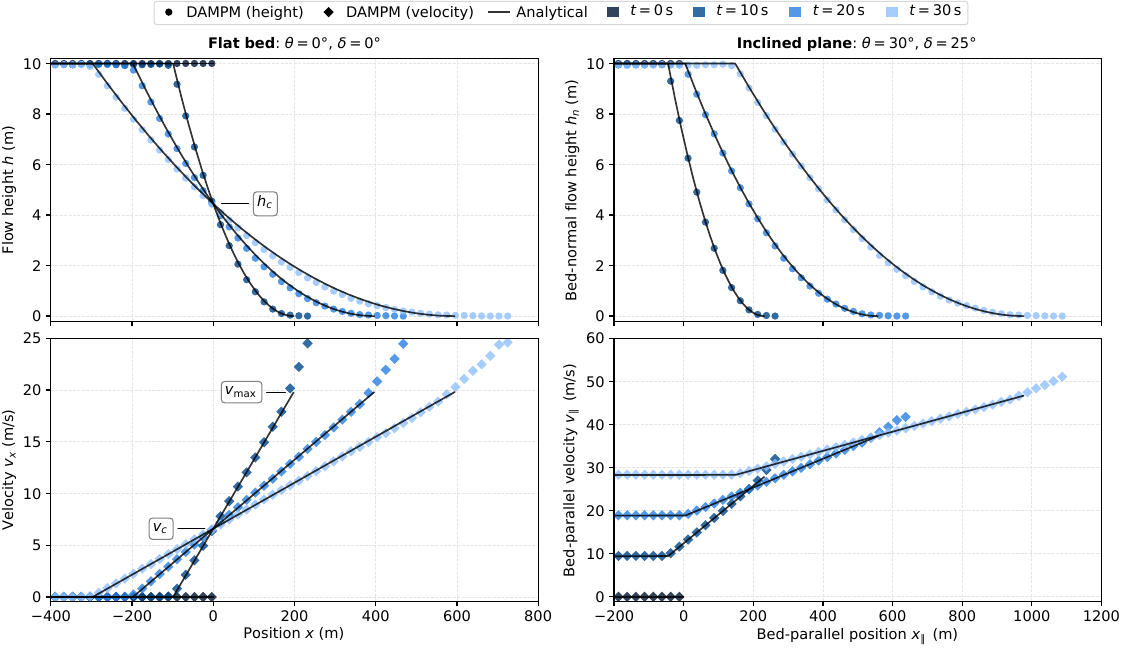}
    \caption{1D dam break over a flat, frictionless bed, and on an inclined plane with Coulomb friction. Depth-averaged MPM results are computed in a bed-aligned coordinate system, simulated from an initial reservoir with a size of \SI{1000}{\meter} $\times$ \SI{200}{\meter}, without cross-slope forces, and averaged over $y$ in bins along $x$. Each plotted data point is the arithmetic mean of all particles in that bin. They are compared with the analytical solution of \citet{mangeney2000analytical}.}
    \label{fig:3_1_2_dam_break_profiles}
\end{figure}

Over a flat, frictionless bed~(Fig.~\ref{fig:3_1_2_dam_break_profiles}, left), the model conserves the reservoir height $h_0$ in the undisturbed upstream region, and the head of the rarefaction wave (upstream corner) is only slightly smoothed. In the subcritical regime~($h > h_c = \frac49 h_0$), the flow height agrees closely with the analytical solution. In the supercritical regime~($h < h_c$), the model slightly underpredicts $h$, with the discrepancy growing over time. The wave front, where $h$ approaches zero, is well captured, although a few particles travel beyond the analytical front position. The velocity agrees closely with the analytical solution in both the subcritical~($v_x < v_c = \frac13 v_\text{max} = \frac23\sqrt{gh_0}$) and supercritical regimes~($v_x > v_c$) and maintains a sharp transition at the upstream corner. The particles beyond the analytical front with near-zero flow height exceed the maximum velocity of the analytical solution.

For the simulation on an inclined plane~(Fig.~\ref{fig:3_1_2_dam_break_profiles}, right), the domain is rotated into a bed-aligned Savage--Hutter coordinate system. The model reproduces $h_n$ and $v_\parallel$ with an accuracy similar to that in the flat case.

\begin{figure}[pos=htbp]
    \centering
    \includegraphics{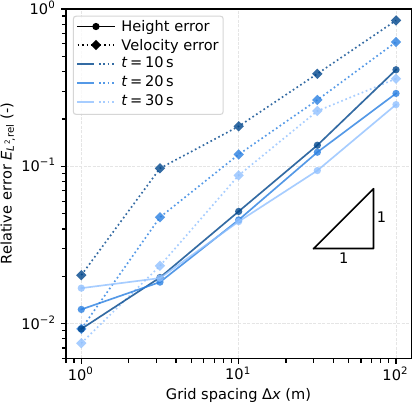}
    \caption{Spatial convergence for the dam break over a flat, frictionless bed~($\theta = \delta = \ang{0}$). Relative $L^2$ error, Eq.~\eqref{eq:L2_error}, of flow height~$h$ and velocity~$v_x$ versus grid spacing~$\Delta x$, evaluated over all bins containing particles between $x = \SI{-400}{\meter}$ and the analytical front position~$x_\text{tip}$. The results are averaged over the \SI{200}{\meter} wide domain in $y$, which comprises only two cells at the coarsest resolution~($\Delta x = \SI{100}{\meter}$).}
    \label{fig:3_1_2_dam_break_convergence}
\end{figure}

We quantify the spatial convergence for the flat, frictionless case with the relative $L^2$ error
\begin{equation}\label{eq:L2_error}
    E_{L^2,\text{rel}}(q) = \sqrt{\frac{\sum_{i \in I_\text{active}} \left(q_{\text{DAMPM},i} - q_\text{ana}(x_i)\right)^2}{\sum_{i \in I_\text{active}} (q_\text{ana}(x_i))^2}}, \qquad q \in \{h, v_x\},
\end{equation}
where $q_{\text{DAMPM},i}$ is the bin average at the bin center $x_i$, and $I_\text{active} = \{i \mid \SI{-400}{\meter} \leq x_i \leq x_\text{tip} \text{ and bin } i \text{ contains particles}\}$ with the analytical front position~$x_\text{tip} = 2\sqrt{g h_0}\,t$. With grid refinement~(Fig.~\ref{fig:3_1_2_dam_break_convergence}), the error decreases roughly linearly.

\subsection{Validation}

We validate our model against the granular column collapse experiment of \citet{lajeunesse2004spreading}, in which a cylindrical granular column with initial radius~$R_0 = \SI{70.5}{\milli\meter}$ and height~$H_0 = \SI{56.4}{\milli\meter}$ spreads on a horizontal plane. In the simulation, the initial condition is an exact cylinder without edge smoothing. We use Coulomb friction with $\mu = 0.61$, calibrated to match the experimental deposit, $\Delta x = \SI{1}{\milli\meter}$, and $\Delta t_\text{max} = \SI{0.1}{\milli\second}$.

\begin{figure}[pos=htbp]
    \centering
    \includegraphics{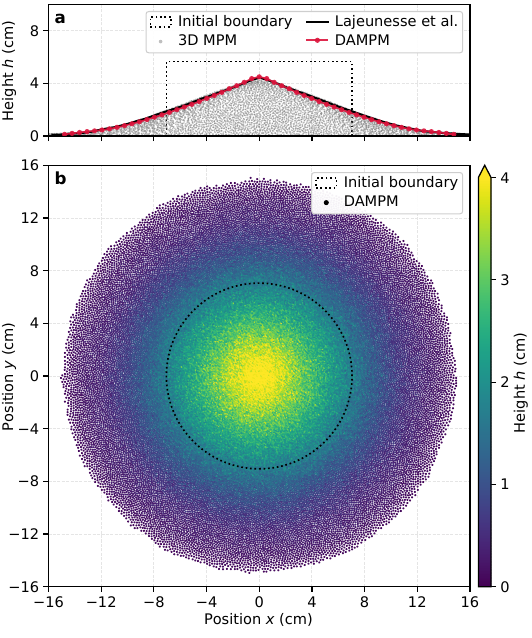}
    \caption{Final deposit of the granular column collapse experiment of \citet{lajeunesse2004spreading}, simulated with depth-averaged MPM using Coulomb friction and \num{62416} particles. \textbf{a}, cross-section profile of the deposit height, compared with the experimental profile and 3D MPM. \textbf{b}, top view of the deposit where each dot represents one particle.}
    \label{fig:3_2_granular_collapse}
\end{figure}

To generate the cross-section profile of the deposit~(Fig.~\ref{fig:3_2_granular_collapse}a), we first calculate the radial distance $r = \sqrt{x^2 + y^2}$ for all particles, bin the data in $r$, and then mirror it. The simulated deposit and the sloping free surface agree well with the experimental profile, which has also been reproduced with 3D MPM~\citep[model of][]{gaume2018dynamic}. The top view shows a circular deposit~(Fig.~\ref{fig:3_2_granular_collapse}b), indicating that the spreading is independent of the grid orientation.

\subsection{Application to 2023 Brienz/Brinzauls Rockslide}\label{sec:brienz}

\begin{figure}[pos=htbp]
    \centering
    \includegraphics{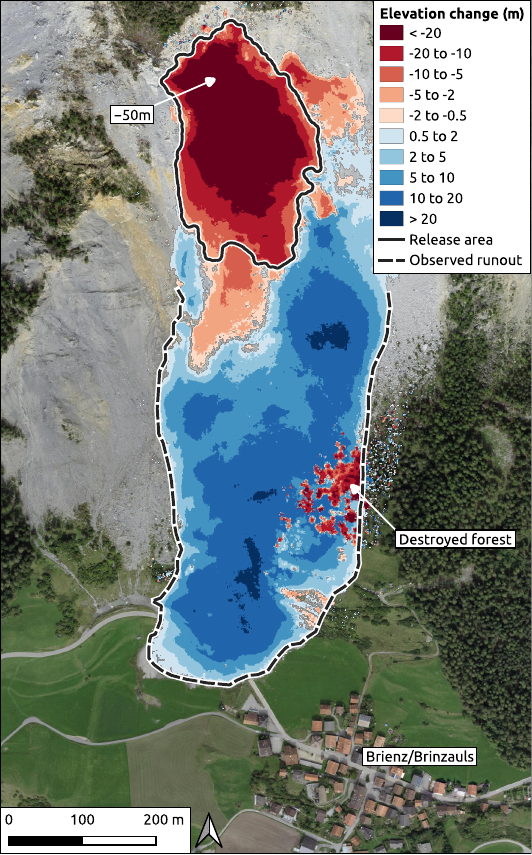}
    \caption{DSM of difference~(DoD) of the 2023 Brienz/Brinzauls rockslide, $\text{DSM}_\text{post} - \text{DSM}_\text{pre}$, computed from drone digital surface models~(DSMs) acquired on June~8, 2023 (pre-event) and June~16, 2023 (post-event). The solid outline delineates the release area for the simulation, in which $\text{DSM}_\text{pre} - \text{DSM}_\text{post} > \SI{2}{\meter}$. The orthophoto shows slightly larger footprint at the toe of the deposit due to settlement. Orthophoto 2025~\textcopyright~swisstopo.}
    \label{fig:3_3_1_study_area}
\end{figure}

On June~15, 2023, a rockslide occurred above the village of Brienz/Brinzauls, Grisons, Switzerland~(Fig.~\ref{fig:3_3_1_study_area}). It originated from the so-called Insel compartment, which had formed within a few years on top of the deep-seated landslide complex on which the village is built~\citep{kenner2025formation, loew2025early}. As the compartment accelerated in early 2023, an early warning system prompted the evacuation of the village in May~\citep{loew2025early}. Following several smaller failures at the toe, most of the mass failed within approximately two minutes, starting at around 23:37 local time~\citep{loew2025early, aaron2026dynamic}. About \SI{1.2e6}{\cubic\meter} of the roughly \SI{2.7e6}{\cubic\meter} compartment left the sliding plane as a dry granular flow that stopped at the edge of the village~\citep{kenner2025formation}.

We use this rockslide event as a real-world benchmark to test the performance of our model. All simulations of this event cover \SI{80}{\second} of physical time, which is when most particles have come to rest. They are run on a desktop computer with an Intel~Core~i9-14900K CPU~(24~cores, 32~threads) and DDR5-5200 dual-channel RAM. Reported simulation times are wall-clock times including I/O, measured from the start of the simulation until all data are saved to disk.

\subsubsection{Back-calculation}\label{sec:brienz_back}

Terrain and release are derived from pre-event and post-event drone digital surface models~(DSMs) with a resolution of \SI{1}{\meter}. The basal topography is $b(\bm x_i) = \min(\text{DSM}_\text{pre}, \text{DSM}_\text{post})$, cleaned by the remaining mass in the release area. The release height is $h_0 = \text{DSM}_\text{pre} - \text{DSM}_\text{post}$, restricted to the area where $h_0$ exceeds \SI{2}{\meter} and excluding those areas that we attribute to entrainment and secondary releases~(Fig.~\ref{fig:3_3_1_study_area}). The basal topography is smoothed by resampling to \SI{5}{\meter} and then interpolated to the simulation resolution of $\Delta x = \SI{1}{\meter}$ with cubic B-splines. We calibrate the friction parameters of each law for the best visual fit of the simulation runout to the observed runout in Fig.~\ref{fig:3_3_1_study_area}, which yields $\mu = \tan\ang{28.5}$ for Coulomb, $\mu = \tan\ang{24.5}$ and $\xi = \SI{1000}{\meter\per\second\squared}$ for Voellmy, and $\mu_1 = \tan\ang{24.5}$, $\mu_2 = \tan\ang{35.5}$, $I_0 = 0.25$, $d = \SI{0.1}{\meter}$, and $\phi = 0.6$ for $\mu(I)$. We use $\Delta t_\text{max} = \SI{0.04}{\second}$.

\begin{figure}[pos=htbp]
    \centering
    \includegraphics[width=\linewidth]{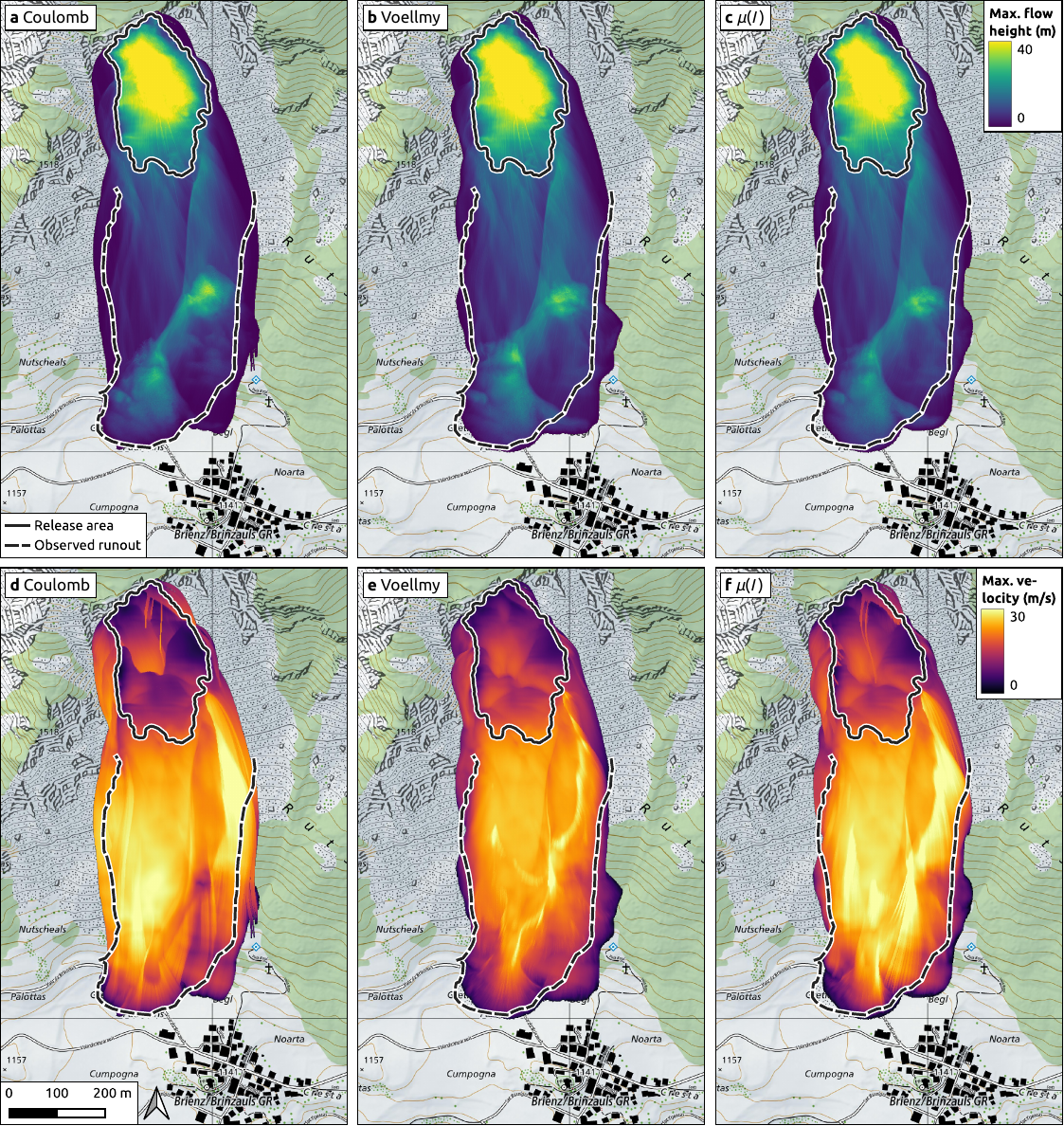}
    \caption{Simulation of the 2023 Brienz/Brinzauls rockslide (simulation resolution $\Delta x = \SI{1}{\meter}$, release volume \SI{1.08e6}{\cubic\meter}, \num{190077} particles). \textbf{a--c}, maximum vertical flow height $h_\text{max}$. \textbf{d--f}, maximum horizontal velocity magnitude~$v_\text{max}$. Solid line: simulation release area. Dashed line: observed runout, which was used to calibrate the simulation parameters. All three simulations (Coulomb, Voellmy and $\mu(I)$) run faster than real time. Topographic map~\textcopyright~swisstopo.}
    \label{fig:3_3_1_hmax_vmax}
\end{figure}

All three friction laws yield very similar maximum flow heights, but the affected area is slightly wider for Coulomb~(Fig.~\ref{fig:3_3_1_hmax_vmax}a--c). Voellmy reaches slightly lower maximum velocities~(Fig.~\ref{fig:3_3_1_hmax_vmax}d--f). Moreover, Voellmy friction yields near-zero velocities at the flow margins, because the turbulent term of the Voellmy law adds a term proportional to $\frac1h$ to the effective friction coefficient. This contribution grows as $h$ decreases and rapidly halts shallow flow.

These single-threaded \SI{1}{\meter} high-resolution simulations take \SI{55}{\second} for Coulomb, \SI{42}{\second} for Voellmy, and \SI{47}{\second} for $\mu(I)$, i.e., they run faster than real time. The runtimes differ because the adaptive time step size depends on the maximum velocity.

\subsubsection{Probability Map}\label{sec:brienz_prob}

\begin{figure}[pos=htbp]
    \centering
    \includegraphics{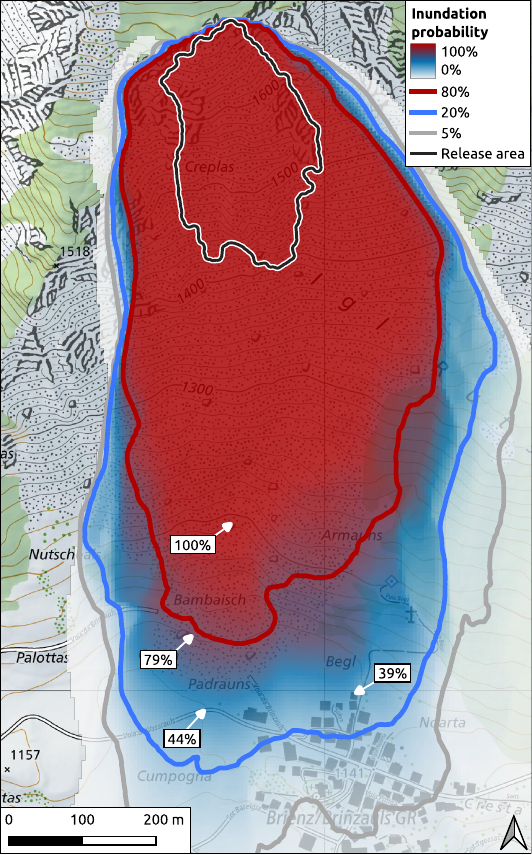}
    \caption{Inundation probability $P(h_\text{max} > \SI{0}{\meter})$, Eq.~\eqref{eq:exceedance_probability}, i.e., the probability that a location is reached by the rockslide, computed from 1632 simulations using Voellmy friction. The parameters $\mu$ and $\xi$ are varied around the back-calculated values $\mu_0 = \tan\ang{24.5}$ and $\xi_0 = \SI{1000}{\meter\per\second\squared}$ with Gaussian weights. Due to the resolution of $\Delta x = \SI{5}{\meter}$~(7664 particles), the entire 1632-simulation ensemble runs in real time, with individual runs being orders of magnitude faster than real time. Topographic map~\textcopyright~swisstopo.}
    \label{fig:probability_map}
\end{figure}

To demonstrate the computational abilities in probabilistic hazard mapping, we run an ensemble of 1632~simulations with Voellmy friction at $\Delta x = \SI{5}{\meter}$ resolution, the friction law and resolution usually used in operational hazard mapping in Switzerland. The parameters are sampled on a regular grid of 51 values of $\mu$ between 0.2 and 0.7, spaced uniformly in linear space, and 32 values of $\xi$ between \SI{200}{\meter\per\second\squared} and \SI{5000}{\meter\per\second\squared}, spaced uniformly in $\log_{10}$ space. We sample $\mu$ more finely because it affects the runout distance more strongly than $\xi$. Moreover, the sampling is logarithmic in $\xi$ because the turbulent term is inversely proportional to it, and because it spans more than one order of magnitude. To account for parameter uncertainty, each simulation~$i$ is assigned the weight~$w_i = w_{\mu,i}\,w_{\xi,i}$, with
\begin{equation}\label{eq:weights}
    w_{\mu,i} = \exp\left(-\frac{(\mu_i - \mu_0)^2}{2\sigma_\mu^2}\right), \qquad
    w_{\xi,i} = \exp\left(-\frac{(\log_{10}(\xi_i) - \log_{10}(\xi_0))^2}{2\sigma_\xi^2}\right),
\end{equation}
centered on the back-calculated parameters $\mu_0 = \tan\ang{24.5} \approx 0.456$ and $\xi_0 = \SI{1000}{\meter\per\second\squared}$, with standard deviations $\sigma_\mu = 0.1$ (in linear space) and $\sigma_\xi = 0.35$ (in $\log_{10}$ space). The Gaussians are truncated at the limits of the sampled parameter ranges. The exceedance probability of a flow height threshold~$h_\text{crit}$ at location~$\bm x$ is
\begin{equation}\label{eq:exceedance_probability}
    P(\bm x) = \frac{\sum_i w_i\,\chi_i(\bm x)}{\sum_i w_i}, \qquad
    \chi_i(\bm x) = \begin{cases} 1, & h_{\text{max},i}(\bm x) > h_\text{crit},\\ 0, & \text{else}, \end{cases}
\end{equation}
where $h_{\text{max},i}$ is the maximum flow height of simulation~$i$. Here, we use $h_\text{crit} = \SI{0}{\meter}$, and perform the simulations with $\Delta t_\text{max} = \SI{0.2}{\second}$. This probability is \SI{100}{\percent} on the main slope and decreases downslope to \SI{79}{\percent} at the first road and \SI{44}{\percent} at the second road~(Fig.~\ref{fig:probability_map}).

As the ensemble is embarrassingly parallel, we execute single-threaded simulations concurrently with GNU Parallel on 32~threads. The 1632~simulations complete in \SI{76}{\second}, i.e., at a throughput of 21.5~simulations per second.

\subsubsection{Computational Performance}

We compare our model's computational performance with the commercial software RAMMS::AVALANCHE~\citep{christen2010ramms} and the open-source solver AvaFrame com1DFA~\citep{tonnel2023avaframe}. We run similar simulations as in Secs.~\ref{sec:brienz_back} and~\ref{sec:brienz_prob}, comparing the performance for a single simulation at \SI{1}{\meter} resolution~(Benchmark~1) and an ensemble of 1000~simulations at \SI{5}{\meter} resolution~(Benchmark~2). We use Voellmy friction with $\mu = \tan\ang{24.5}$ and $\xi = \SI{1000}{\meter\per\second\squared}$ in Benchmark~1, and perform a parameter sweep in Benchmark~2 with 40~values of~$\mu$ between 0.2 and 0.7, and 25~values of~$\xi$ between \SI{200}{\meter\per\second\squared} and \SI{5000}{\meter\per\second\squared}, both spaced uniformly in linear space. As RAMMS does not support spatially variable release heights, we choose a uniform release height for all three models that achieves the same volume as in Sec.~\ref{sec:brienz_back}: $h_0 = \SI{22.59}{\meter}$ in our model (vertical), and $d_0 = \SI{16.62}{\meter}$ in RAMMS and com1DFA (bed-normal). All simulations run for \SI{80}{\second}. Stopping criteria are disabled. Benchmark~1 runs on one thread, and Benchmark~2 on 32~concurrent threads with one simulation per thread. In our model, we choose $\Delta t_\text{max}$ as in Secs.~\ref{sec:brienz_back} and~\ref{sec:brienz_prob}. By choosing \verb+deltaTh=5.4m+ in com1DFA, the number of particles per cell is reduced to 4.12 at \SI{1}{\meter} and 4.01 at \SI{5}{\meter} for comparability with our model, which uses 4~particles per cell. This gives \num{200505} particles in com1DFA and \num{191817} in our model in Benchmark~1, and 8279 and 7700 in Benchmark~2, respectively. The difference probably stems from how particles are created at the release area boundary. In com1DFA, we set the SPH kernel radius to $r_\text{kernel}=\SI{5}{\meter}$ in both benchmarks. Otherwise, all models run with default numerical settings. None of the models save time-step data during these benchmarks.

\begin{figure}[pos=htbp]
    \centering
    \includegraphics{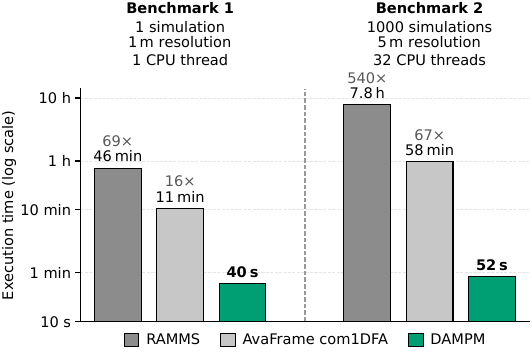}
    \caption{Simulation time in RAMMS::AVALANCHE~2.1.08~\citep{christen2010ramms}, AvaFrame com1DFA~2.2b1~\citep{tonnel2023avaframe}, and our model. Numbers above bars give execution time and factor relative to our model. The simulations are run on a desktop computer whose hardware is given in Sec.~\ref{sec:brienz}.}
    \label{fig:3_3_3_performance}
\end{figure}

We run each scenario twice with each software and report the shorter runtime~(Fig.~\ref{fig:3_3_3_performance}). RAMMS reproducibly crashes during export of the raster data in Benchmark~2, so that the simulation results are inaccessible and, unlike for the other models, its runtime excludes the data output. The solver com1DFA is numerically unstable in Benchmark~1, but not with a reduced release height. Presumably, the gradient at the edge of the release area becomes too large at \SI{1}{\meter} resolution, producing a large lateral pressure. Nevertheless, the measured runtime should accurately reflect the computational performance of com1DFA. Our model is 69~times faster than RAMMS and 16~times faster than com1DFA in Benchmark~1, and 540 and 67~times faster in Benchmark~2.

\section{Discussion}\label{sec:discussion}

In the rigid-block tests, our method reproduces the correct transient and steady-state motion~(Figs.~\ref{fig:3_1_1_velocity_vs_position} and~\ref{fig:3_1_1_steady-state_velocity}). The slight underestimation of $v_\infty$ at steep slopes results from the semi-implicit treatment of the rate-dependent friction laws. Because $\mu_\text{eff}$ is evaluated with the trial velocity~$\bm v_i^*$, the steady state is reached when the trial velocity, rather than the end-of-step velocity, equals $v_\infty$. The 3D end-of-step velocity (in bed-parallel direction) therefore remains below $v_\infty$ by the frictionless velocity increment of one time step, $\Delta t\,g\sin\theta$. This deficit is small, e.g., \SI{0.2}{\meter\per\second} for $\Delta t = \SI{0.04}{\second}$ and $\theta = \ang{30}$. Evaluating $\mu_\text{eff}$ with the end-of-step velocity using an implicit scheme would recover $v_\infty$ exactly. We use the semi-implicit scheme for efficiency and because the relative error is only on the order of \SI{0.5}{\percent}.

The dam-break tests~(Fig.~\ref{fig:3_1_2_dam_break_profiles}) verify the pressure gradient term. Our method reproduces the speeds of the rarefaction wave's upstream-propagating head and of the downstream wet--dry front. True 1D boundary conditions could further improve accuracy and allow for 1D execution instead of $y$-averaging.

The sloping free surface in the granular column collapse~(Fig.~\ref{fig:3_2_granular_collapse}) matches the experiment, and the circular deposit shows no grid-orientation bias. \citet{mangeney2005use} simulated the same configuration with a depth-averaged finite volume model in which friction is treated as an apparent topography. On a horizontal bed, both models solve the same Saint--Venant equations with isotropic stress and Coulomb friction, but with fundamentally different numerical schemes that both bring the mass to rest once the driving force falls below the Coulomb threshold, without a stopping criterion. Both reproduce the experimental deposit, and their friction angle~($\delta = \ang{32}$), calibrated on a smaller aspect ratio, differs by only \ang{0.6} from ours~($\mu = 0.61 \approx \tan\ang{31.4}$). This agreement between independent numerical methods further increases confidence in our model. Both values exceed the repose and avalanche angles of the beads~($\theta_r \approx \ang{21}$, $\theta_a \approx \ang{29}$) because the calibrated friction coefficient is an effective parameter: with the isotropic stress closure~($K = 1$), basal friction must compensate for the missing internal shear resistance. Although the initial aspect ratio violates the shallowness assumption, \citet{mangeney2005use} found that their model reproduces the deposit and the spreading dynamics until the arrest of the front for $a = H_0/R_0 \lesssim 1$, as the column rapidly flattens and spreads through surface avalanches with small vertical accelerations. For $a = 5.4$, the simulated spreading was much faster than in the experiment, and only the deposit was approximately captured. Their model did not capture the arrest phase (for $a = 0.8$, the simulated arrest started in the center and propagated outward, whereas it started at the front in the experiment), which they related to a vertical propagation of the static/flowing interface that depth-averaged models cannot resolve. Our model presumably shares this limitation.

The friction algorithm transitions flowing material to rest when driving shear stresses fall below static friction, preventing residual creep common in other models~\citep{christen2010ramms, rauter2016snow, goodwin2024has}. The $\mu(I)$-rheology assumes a Bagnold profile for velocity, whereas the depth-averaged momentum conservation equation, Eq.~\eqref{eq:momentumBalance}, assumes a uniform velocity profile. The two profiles differ in the momentum shape factor~$\beta = \overline{u^2}/\overline u^2$ (overbars denote depth averages): $\beta = 1$ for a uniform profile and $\beta = 5/4$ for the Bagnold profile~\citep{gray2014depth}. Like most models, our $\mu(I)$ implementation thus combines a basal resistance derived for a sheared flow with the momentum transport of a plug flow. Furthermore, although the $\mu(I)$ parameters can be measured in the laboratory, at field scale, where grain sizes span several orders of magnitude, they have to be calibrated with care, like $\mu$ and $\xi$ of the Voellmy law.

Eulerian FVM models evolve flow height and momentum as conserved quantities across cell faces. Depth-averaged MPM updates the particle flow height from the divergence of grid velocity, which might lead to inferior local conservation properties. Tracking the wet/dry boundary with $h\to0$ is notoriously difficult in FVM models and can cause numerical instability, whereas depth-averaged MPM handles the moving flow margin naturally via Lagrangian particles. As flow height is advected with the particles, it is not smeared by the numerical diffusion of flux-based schemes~\citep{leveque2002finite}. Instead, the particle--grid transfers introduce interpolation errors and, with PIC, numerical dissipation. Compared with SPH, MPM avoids neighbor searches, and the PIC/FLIP blending instead of artificial viscosity controls the trade-off between numerical dissipation and stability.

Unlike previous depth-averaged MPM implementations for granular flows, our model is open-source and designed for large-scale probabilistic hazard assessment. Its speedup over RAMMS~\citep{christen2010ramms} and AvaFrame com1DFA~\citep{tonnel2023avaframe}, two of the most established models in hazard assessment, grows from 16--69 for a single simulation to 67--540 for ensembles~(Fig.~\ref{fig:3_3_3_performance}), where throughput is decisive. From the single \SI{1}{\meter} run to the \SI{5}{\meter} ensemble, the per-simulation time of RAMMS drops by a factor of 98 (from \SI{46}{\minute} to \SI{28}{\second}), whereas the coarser resolution alone would ideally yield a factor of 125, further multiplied by the gain from running 32~simulations concurrently. For com1DFA and our model, this factor is about 190 and 770, respectively. The small gain of RAMMS points to a large per-run overhead (initialization, I/O) and poor parallel efficiency. RAMMS::LSHIM~\citep{buhler2022automated}, a version without graphical user interface designed for large-scale mapping, may behave differently from the standard version tested here.

In an additional test, our single-threaded \SI{1}{\meter} simulation runs about as fast as RAMMS on a high-end consumer GPU, indicating a much better performance per hardware cost. As runtimes depend on model configuration, spatial resolution, and hardware, these comparisons indicate orders of magnitude rather than a precise benchmark, and comparisons with further models are needed to generalize them. Moreover, RAMMS and com1DFA offer several functionalities that are not yet implemented in our model.

Large-scale hazard indication mapping involves millions of runout simulations at \SI{10}{\meter} resolution, e.g., 2.0~million with RAMMS::LSHIM on a 96-core server in about 10~days~\citep{buhler2022automated}, 1.5~million with MoT-Voellmy on five workstations with 64~cores in total in 2--3~weeks~\citep{issler2023naksin}, and 3.5~million with RAMMS::LSHIM on a 184-core HTCondor cluster~\citep{fischer2026automated}. The release areas in such studies are typically much smaller than at Brienz, and the simulations shorter. Extrapolated from Benchmark~2, a study with $10^6$--$10^7$ runouts would take hours to about half a day with our model on a single desktop computer, days to weeks with com1DFA, and months to almost a year with RAMMS, depending on grid resolution. Since energy consumption scales roughly with computation time, the energy demand of such studies may decrease by similar factors.

Owing to the efficiency of our method, our back-calculation of the Brienz/Brinzauls rockslide uses \num{190077}~particles at \SI{1}{\meter} resolution~(Fig.~\ref{fig:3_3_1_hmax_vmax}), whereas \citet{aaron2026dynamic} used 4000~particles at \SI{5}{\meter} resolution. Their results show artificial deposits more than \SI{50}{\meter} thick at the margins of the release zone, a numerical artifact of their interpolation algorithm, which is not present in our method. Their release volume of \SI{2.1e6}{\cubic\meter}, extrapolated from a 3D rupture surface, is about twice ours~(\SI{1.08e6}{\cubic\meter}) and includes material that remained in the source zone, which in our DSM of difference~(DoD)-based setup is part of the basal topography. This difference also affects the calibrated friction parameters: \citet{aaron2026dynamic} obtained $\mu=\tan\ang{29}$ and $\xi = \SI{4600}{\meter\per\second\squared}$, whereas we obtain $\mu=\tan\ang{24.5}$ and $\xi = \SI{1000}{\meter\per\second\squared}$. Their higher $\mu$ is consistent with the considerably larger mass retained in their release area. With $\xi = \SI{4600}{\meter\per\second\squared}$, the turbulent term is almost negligible, so their runout is controlled mainly by $\mu$, whereas our lower $\mu$ is compensated by stronger turbulent friction. That two different parameter sets reproduce the observed runout illustrates the equifinality of the Voellmy model: deposit data alone cannot constrain both $\mu$ and $\xi$. Velocity data, e.g., from acoustic, radar, or seismic records, are needed to separate them. Our single-stage release reproduces the observed runout, consistent with the conclusion of~\citet{aaron2026dynamic} that the main mass failed in one stage within about \SI{2}{\minute}.

A limitation of our setup is that the DoD integrates all surface changes between the two acquisitions, including deformation before and after the event. Compared with the observed deposit~(Fig.~\ref{fig:3_3_1_study_area}), our simulation~(Fig.~\ref{fig:3_3_1_hmax_vmax}) slightly underpredicts the deposit height at the toe and produces a wider flow. In the real event, the forest in the east~(Fig.~\ref{fig:3_3_1_study_area}) influenced the flow. As the model neglects forest, the simulated flow continues into this area and deposits material east of the observed deposit.

The probability map~(Fig.~\ref{fig:probability_map}) demonstrates how our method can be used in probabilistic hazard assessment. The inundation probabilities only account for the uncertainty of the Voellmy parameters. As the ensemble completes within \SI{76}{\second}, it can be extended to uncertainties in release volume and area, and with weights derived from a calibration against documented events.

Several limitations of this model can be addressed in further work. Curvature effects can be included as a centripetal contribution to the basal normal stress which increases friction in concave terrain~\citep{pudasaini2003rapid, peruzzetto2021topography}. Entrainment, which can substantially increase the volume and mobility of mass flows~\citep{iverson2015entrainment}, can be represented by increasing particle masses or inserting particles. The isotropic stress closure can be extended to active and passive stress states~\citep{savage1989motion} or general elastoplastic and viscoplastic models~\citep{guillet2023depth, meloche2025modeling}. A momentum shape factor $\beta \neq 1$ can be included in the momentum conservation equation. Pore pressure, whose effect on debris-flow mobility can currently only be represented through effective friction parameters, can be incorporated in a two-phase formulation~\citep{pastor2009depth,iverson2014depth,mergili2017r, juel2026stabilized}. Parallelization of individual simulations, potentially on GPUs, would accelerate large high-resolution simulations~\citep{zhao2026geowarp}. Finally, further validation should address other processes, such as snow avalanches and debris flows, relying on remote sensing data, seismic signals, or in situ measurements.

\section{Conclusion}\label{sec:conclusion}

We presented an efficient open-source depth-averaged material point method for geophysical mass flows. It reproduced analytical solutions for a sliding block and for dam-break problems and matched the deposit of a laboratory granular column collapse without grid-orientation bias. The separate friction step brought the flow to rest without artificial stopping criteria.

The back-calculation of the 2023 Brienz/Brinzauls rockslide at \SI{1}{\meter} resolution ran faster than real time on a single CPU core. After calibration to the observed runout, Coulomb, Voellmy, and $\mu(I)$ friction yielded very similar results, but Coulomb friction showed a slightly wider affected area. Runout observations alone are therefore insufficient to discriminate between friction laws. Our implementation reached a throughput of more than 20~simulations per second at \SI{5}{\meter} resolution and was 16 to 69~times faster than AvaFrame com1DFA and RAMMS for a single simulation and 67 to 540~times faster for an ensemble. Probabilistic hazard assessment, sampling uncertainties in friction parameters, release area, and release volume, thus becomes feasible on a desktop computer, even for large-scale hazard indication mapping with millions of simulations.

\section*{Acknowledgements}

We gratefully acknowledge N.~Oestreicher and Y.~Bühler of the Alpine Remote Sensing group, WSL Institute for Snow and Avalanche Research SLF, Davos, for providing processed drone data of the 2023 Brienz/Brinzauls rockslide. We thank D.~Steffen and L.~Guillet for fruitful discussions.

\section*{Declaration of Competing Interests}

The authors declare that they have no known competing financial interests or personal relationships that could have appeared to influence the work reported in this paper.

\section*{Funding}

Yidong Zhao acknowledges financial support from the Swiss National Science Foundation~(SNSF) through grant number CRSK-2\_237763. Lars Blatny acknowledges financial support from the Swiss National Science Foundation~(SNSF) through grant number P500PT\_230265.

\section*{Declaration of Generative AI Use}

During the preparation of this work the authors used Gemini and Claude to assist with the writing. After using these tools, the authors reviewed and edited the content as needed and take full responsibility for the content of the published article.

\section*{Data Availability Statement}

The source code and the DEM data are available on GitHub (\href{https://github.com/ALMO-SLF-ETH/Slab-Pro}{github.com/ALMO-SLF-ETH/Slab-Pro}). Release v0.1.0 was used in this study.

\bibliographystyle{cas-model2-names}
\bibliography{references}

\end{document}